# CO$_2$-induced Drastic Decharging of Dielectric Surfaces in Aqueous Suspensions

Peter Vogel[1]*, David Beyer[2], Christian Holm[2], and Thomas Palberg[1]

[1] Institute of Physics, Johannes Gutenberg University, 55128 Mainz, Germany

[2] Institut für Computerphysik (ICP), Universität Stuttgart, 70569 Stuttgart, Germany

* corresponding author e-mail: petvogel@uni-mainz.de

**Abstract:** We study the influence of airborne CO$_2$ on the charge state of carboxylate stabilized polymer latex particles suspended in aqueous electrolytes. We combine conductometric experiments interpreted in terms of Hessinger's conductivity model with Poisson-Boltzmann cell (PBC) model calculations with charge regulation boundary conditions. Without CO$_2$, a minority of the weakly acidic surface groups are dissociated and only a fraction of the total number of counter-ions actually contribute to conductivity. The remaining counter-ions exchange freely with added other ions like Na$^+$, K$^+$ or Cs$^+$. From the PBC-calculations we infer a corresponding pK$_a$ of 4.26 as well as a renormalized charge in reasonably good agreement with the number of freely mobile counter-ions. Equilibration of salt- and CO$_2$-free suspensions against ambient air leads to a drastic de-charging, which exceeds by far the expected effects of to dissolved CO$_2$ and its dissociation products. Further, no counter-ion-exchange is observed. To reproduce the experimental findings, we have to assume an effective pK$_a$ of 6.48. This direct influence of CO$_2$ on the state of surface group dissociation explains our recent finding of a CO$_2$-induced decrease of the $\zeta$-potential and supports the suggestion of an additional charge regulation caused by molecular CO$_2$. Given the importance of charged surfaces in contact with aqueous electrolytes, we anticipate that our observations bear substantial theoretical challenges and important implications for applications ranging from desalination to bio-membranes.

## 1. Introduction

Dielectric surfaces in contact with aqueous electrolytes may charge up by dissociation of ionogenic surface groups or by adsorption of ions, and develop an electric double layer (EDL) by counterion accumulation [1]. Due to pH-driven charge regulation [2], this process depends on the surface chemistry and the pH of the adjacent electrolyte solution [3,4,5]. Further, the type of added electrolytes and its concentration influence the charge state [6,7,8,9]. In addition, it varies with the degree of double layer overlap between suspended dielectric particles [10,11,12,13]. Constituting a many-body soft matter system, the charging process and the structure of the EDL are of fundamental interest but at the same time are quite complex [1,2]. Approaches on different levels of sophistication have been developed for theoretical calculations of the charge state of dielectric surfaces [14,15,16,17,18,19,20,21,22] as well as for the determination of electro-kinetic properties and their interpretation [23,24,25,26,27,28,29,30,31,32]. Moreover, charging bears important implications for the functioning of dielectric surfaces in many contexts ranging from biology over climate physics to waste water treatment [33,34,35,36,37].

A particularly interesting class of potentially charge-regulating additives are dissolved molecular gases like $O_2$, $N_2$ and $CO_2$. These are inevitably present in any system allowed to have contact with ambient air [38,39] including standard doubly distilled water or dialyzed systems, but can be efficiently removed by degassing. Molecular gases may adsorb reversibly on dielectric surfaces [40,41], alter the surface structure [42], and influence the colloid stability [43,44]. Unlike its inert counterparts, $CO_2$ partially dissociates to form carbonic acid [45,38,39], which in turn provides additional screening electrolyte and shifts the pH [46]. Increasing the electrolyte content affects double layer extension leading to a drop in the Debye screening length. At the same time, a decrease in bulk pH affects the charge state by a shift in the surface group dissociation equilibrium [47,48,49] thus resulting in altered suspension properties [50,51]. It therefore is paramount to work at well-defined $CO_2$-concentrations [52], e.g., by initial selective $CO_2$-removal *via* exhaustive ion exchange in air-tight vessels or closed conditioning circuits and subsequent addition *via* contact with ambient air [53,54,23].

Recently, we performed electrophoretic experiments at constant background concentrations of the inert gases but various concentrations of molecular $CO_2$ [55]. There, we found the effects of molecular $CO_2$ on the charge state of dielectric surfaces to go way beyond the expected effects of screening and pH-variation [10,47,48,56]. In particular, upon equilibration in contact with ambient air, the magnitudes of the electrokinetic $\zeta$-potential dropped significantly. The process could be driven even further towards full discharging by equilibrating the suspension against pure $CO_2$-atmospheres [57]. Accompanying MD simulations revealed the formation of a thin, diffuse layer of molecular $CO_2$ very close to the surface for both hydrophobic and hydrophilic dielectric materials. From that, we proposed *dielectric* charge regulation as the underlying mechanism [55]: The presence of $CO_2$ lowers the local dielectric constant and thus enhances the electrostatic interaction at and/or between dissociated surface groups. The corresponding Coulomb-energy causes an increase of the dissociation energy and/or introduces an energetic penalty for closely spaced surface groups. In effect this leads to a decrease of the $\zeta$-potential. However, no direct information on the degree of dissociation, the bare charge, the effective charge or the surface chemistry is obtained from electrophoretic experiments.

The present study therefore complements our previous investigations by a comprehensive conductometric charge characterization as well as by Poisson-Boltzmann cell (PBC) model calculations with charge regulation boundary conditions. As a model system, we use negatively charged polystyrene particles of low size dispersity carrying ionizable carboxylate groups from synthesis. Specifically, we perform conductivity experiments in dependence on different variables, like number density, $n$, or type and concentration $c$ of added 1:1 electrolyte (NaCl, HCl, NaOH). Our results are interpreted within Hessinger´s model of independent ion migration [58]. The model gives access to three complementary charge numbers: the number of ionogenic surface groups $N$, the bare charge, $Z$, and the number of freely mobile counterions, i.e., the effective conductivity charge $Z_\sigma$. It successfully described the density and temperature dependent conductivity of deionized and decarbonized suspensions [59], it applies to single component suspensions as well as mixtures thereof [54]. While a rigorous theoretical description of the relation between different effective charges is still under discussion [60,61,62] experimentally determined effective conductivity charges generally agree rather well with effective electrokinetic charges and with renormalized charges, at least as long highly charged particles in 1:1 electrolytes are considered [8,25].

We compare experiments in the absence and presence of $CO_2$ to isolate and discuss the effects of carbonization. We show that the presence of $CO_2$ has significant quantitative and even qualitative effects on the resulting density- and concentration-dependent conductivities which go way beyond the effects expected for dissolved $CO_2$ and its dissociation products. Next to an overall $CO_2$-induced discharging, manifest via decreased bare and effective charges, we observe the absence of ion-exchange between the immobile and mobile EDL populations. This unexpected finding is shown by the PBC-calculations to relate to a significant increase of the effective surface group $pK_a$. Our results therefore confirm and extend the results of our previous electrophoretic experiments in several ways. We anticipate that our findings will stir further theoretical interest in the structure and properties of the EDL at charged dielectric surfaces under ambient conditions and may contribute to a fundamentally improved understanding of the resulting performance of dielectric surfaces.

In what follows, we first shortly introduce the experimental system and our conditioning procedures techniques before describing the actual conductivity measurements and recalling Hessinger's conductivity model. The results section shows the obtained conductivities and interprets these in terms of characteristic charge numbers. We rationalize our findings using complementary Poisson-Boltzmann cell model calculations. We then discuss our findings in relation to the recently proposed decharging of dielectric surfaces by adsorbed molecular $CO_2$. We conclude with a short outlook on further experiments and some practical consequences of our results.

## 2. Materials and sample preparation

We used spherical copolymer particles synthesized by radical emulsion-polymerization from 60% w/w butylacrylamid, 38% w/w styrene, 2% w/w acrylic acid, and 0,5% w/w Sodium-peroxodisulfate, kindly provided by BASF, Ludwigshafen (Lab code PNBAPS359, manufacturer batch number 2168/7390). Their diameter of $2a = 359$nm was determined by the manufacturer using analytical ultracentrifugation. Stock suspensions of particle number densities $n \approx 1 \times 10^{18}$m$^{-3}$ were prepared by dilution with deionized (DI) water. Subsequently, mixed-bed ion exchange resin (IEX) (Amberjet, Carl Roth GmbH + Co. KG, Karlsruhe, Germany) was added and the suspensions were left to stand under occasional stirring for some weeks. They were then coarsely filtered using Sartorius 5µm syringe-driven membrane filters to remove dust, ion-exchange debris, and coagulate regularly occurring upon first contact with IEX. All further sample preparation/conditioning steps including the conductivity measurements were performed in a peristaltically driven circuit described in detail elsewhere [54].

Briefly, appropriately diluted stock suspensions were loaded into the circuit under filtering with Sartorius 1.2 µm and 0.8µm filters. Then, the suspension is pumped through the air-tight tubings connecting (i) the IEX filled ion exchange chamber, (ii) a reservoir under inert gas atmosphere (Ar) to add suspension, water, or electrolyte solutions, (iii) the conductivity measurement cell and (iv) a rectangular quartz-cell (5mm × 10mm) for turbidity measurements [52] for *in situ* control of the number density, $n = 3\Phi/(4\pi a^3)$, respectively the particle volume fraction, $\Phi$.

## 2. Conductivity experiments

Bulk conductivities, $\sigma$, were measured at a frequency of $\omega = 400$Hz (electrodes LTA01 and LR325/01 with bridge LF538 or electrode LR325/001 with bridge LF340, WTW, Germany). This frequency is low enough to stay in the DC-limit, but large enough to avoid electrode polarization effects.

For measurements starting from the thoroughly deionized and decarbonized reference state, the sample is cycled until after about an hour a stable minimum conductivity is reached. Control experiments with pure water yield a background conductivity of $\sigma_B$ = (55-60)nScm$^{-1}$ due to auto-dissociation of water. For experiments in presence of CO$_2$, thoroughly deionized and decarbonized suspensions are brought into contact with ambient air by opening the reservoir while cycling under bypass of the IEX column. After about half an hour, the CO$_2$ dissolving and dissociation reactions have equilibrated increasing $\sigma_B$ by the amount of carbonic acid present in solution. In the electrolyte concentration dependent measurements, starting from either state, the IEX column was bypassed and electrolyte solution (Titrisol 0.1

molL⁻¹ NaCl, HCl or NaOH, Merck, Germany) was added in small quantities under conductometric control.

Measured conductivities are interpreted using Hessinger´s conductivity model and reported in terms of charge numbers. This model approximates the radial dependence of micro-ionic diffusivities [12] by considering a bimodal distribution of either zero or bulk mobility for the counter-ions. The composition of both fractions is assumed to be identical. Assuming Drude-type conductivity, the model sums over the conductivity contributions of all ionic species $i$:

$$\sigma = \sum_i n_i \, e \, z_i \, \mu_i \tag{1}$$

Here, $e$ is the elementary charge, $n_i$ is the number density, $\mu_i$ the mobility, and $z_i$ the valency of any species, respectively. Note that the sum includes the contributions of particles, counter-ions, ions from water auto-protolysis, added (and reacted) electrolyte ions as well as the dissociation products of $CO_2$. The respective micro-ionic mobilities $\mu_i$ are taken from literature [63], while the particle mobilities $\mu_P$ are determined in independent experiments [52]. To be specific, we consider a 1:1 electrolyte at $T = 298.15$ K, where the valence is $z_i = 1$, and the number density of each micro-ionic species is $n_i = c_i 1000 N_A$, with $c_i$ being the electrolyte molar concentration. At ambient conditions, the saturation concentration of ambient $CO_2$ dissolved in water amounts to $c_{CO2} = 1.18 \times 10^{-5}$ molL⁻¹ ($n_{CO2} = 7.1 \times 10^{21}$ m⁻³). Due to partial dissociation of its reaction product carbonic acid, $c_{HCO3-} = c_{H+} = 3.16 \mu$molL⁻¹, and we have pH = 5.5 [46]. Further, at 298.15K, the bulk mobilities of particles, H⁺, OH⁻, CO₃H⁻, Na⁺, and Cl⁻ are $\mu_P \approx (2-5) \times 10^{-8}$ m²s⁻¹V⁻¹ [52], $\mu_{H+} = 36.23 \times 10^{-8}$ m²s⁻¹V⁻¹, $\mu_{OH-} = 20.52 \times 10^{-8}$ m²s⁻¹V⁻¹, $\mu_{HCO3-} = 4.61 \times 10^{-8}$ m²s⁻¹V⁻¹, $\mu_{Na+} = 5.19 \times 10^{-8}$ m²s⁻¹V⁻¹, and $\mu_{Cl-} = 7.91 \times 10^{-8}$ m²s⁻¹V⁻¹, respectively [63].

This yields a general expression to describe the conductivity of suspensions containing negatively charged particles with their counter-ions and additional non-reacting 1:1 electrolyte:

$$\sigma = ne\left(\mu_P + \langle \mu^+ \rangle\right) + M\left(\langle \mu^+ \rangle + \langle \mu^- \rangle\right) + \sigma_B \tag{2}$$

Here, we used the arithmetic means of all cationic (+), respectively anionic (−) micro-ionic mobilities:

$$\langle \mu^+ \rangle = \frac{\sum_i \mu_i M_i^+}{\sum_i M_i^+} \tag{3a}$$

$$\langle \mu^- \rangle = \frac{\sum_i \mu_i M_i^-}{\sum_i M_i^-} \tag{3b}$$

including the $Z$ protonic counter-ions in Eqn (3a). We further introduced the number of added micro-ions of species $i$ per particle, $M_i = n_i/n = c\ 1000\ N_A / n$. Their total number then is $M = \Sigma M_i$. The background concentration of micro-ions, $\sigma_B$, is composed of contributions of ions from water auto-dissociation, from residual ions e.g., leaking from the container walls, and from partially dissociating $CO_2$. In deionized, decarbonized water, the contribution to the micro-ionic concentration from auto-dissociation is $c \approx 2\times10^{-7}$ MolL$^{-1}$, yielding $\sigma_{H2O} = 55$nScm$^{-1}$. This contribution is important in systems of very small $n$ [52]. The level of unidentified small ions typically does not exceed $c \approx 10^{-8}$ MolL$^{-1}$ [64], and its conductivity contribution can usually be neglected. However, suspensions equilibrated against ambient air contain small amounts of carbonic acid, and we use $\sigma_{CO3H2} = 3.16\mu$MolL$^{-1}$ 1000 $N_A$ e ($\mu_{H+} + \mu_{HCO3-}$).

## 3. Experimental results

*3.1. Particle number density dependent conductivities in the deionized state*

Figure 1 compares the particle number density dependence of the background-corrected conductivity of deionized suspensions in the decarbonized state and the $CO_2$-equilibrated state. In both cases, we obtain a strictly linear dependence of the background corrected conductivity on the particle number density (dashed lines). However, in the presence of ambient $CO_2$, we observed a significantly lower slope. For evaluation, we rewrite Eqn (2) to read:

$$\sigma = \sigma_0 + \sigma_B = neZ_\sigma(\mu_P + \mu_{H^+}) + \sigma_B. \tag{4}$$

Least squares fits of Eqn (4) to the data in Fig. 1 (dashed lines) yield $Z_\sigma = (2.35 \pm 0.04)\times10^3$ for the decarbonized and $Z_\sigma = (1.30 \pm 0.07)\times10^3$ for the $CO_2$-saturated state. Where not indicated otherwise, quoted errors denote the standard error at a confidence level of 0.95. The corresponding diffuse layer charge-densities are: $|\sigma_{el}| = (9.3 \pm 0.1)\times10^{-2}$ µCcm$^{-2}$ and $|\sigma_{el}| = (5.1 \pm 0.1)\times10^{-2}$ µCcm$^{-2}$, respectively, where $|\sigma_{el}| = Z_\sigma e/4\pi a^2$ [37]. We thus see a 45% reduction of the conductivity charge in the presence of $CO_2$.

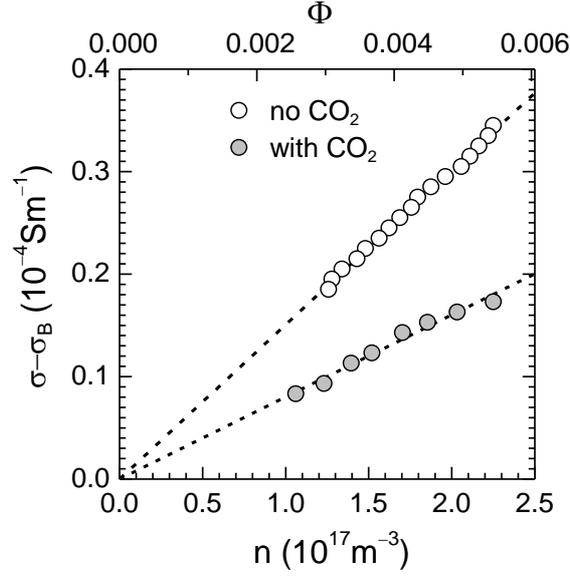

**Figure 1. Density-dependent conductivity of deionized colloidal suspensions.** Background-corrected conductivities as a function of number density $n$ for the two representative states as indicated in the key. The upper scale shows the volume fraction, $\Phi = (4/3)\pi a^3 n$ with $2a = 359$nm being the particle diameter. The dashed lines are least squares fits of Eqn (7) returning values of $Z_\sigma = (2.35 \pm 0.04) \times 10^3$ and $Z_\sigma = (1.30 \pm 0.07) \times 10^3$ for the $CO_2$-free and the $CO_2$-saturated cases, respectively.

*3.2 Conductivity titration with NaOH*

Upon adding a reactive species like NaOH to thoroughly deionized and decarbonized suspensions, we observe the well-known non-monotonic conductivity dependence in Fig. 2a [3,5]. Each background-corrected data set shows an initial, roughly linear descent. Past the minimum we observe a moderate, again roughly linear ascent, which steepens past the equivalence point. For increasing $n$, the minima shift upward and further to the right. In Fig. 2b, we replot the data versus $M^+$. Now the minima are at the same salt concentration. From fits of linear functions to the middle and final parts we infer the total number of surface groups to be $N = (134 \pm 2) \times 10^3$. During the initial, $n$-dependent decrease of conductivity, only dissociated ions are exchanged. Equation (2) reads:

$$\sigma - \sigma_B = \sigma_0 + ne\Delta\mu M^+ \frac{Z_\sigma}{Z} \qquad (5)$$

Here the mobility difference $\Delta\mu = \mu_{Na+} - \mu_{H+}$ has been introduced. The initial slope is $d\sigma/dM^+ = ne\Delta\mu Z_\sigma/Z$. With $Z_\sigma$ taken from the density dependent experiment at no added salt, fits of Eqn (5) to the initial descent yield the charge ratio $Z_\sigma/Z$, respectively the bare charge $Z$. From the fits shown in Fig. 2b as red solid lines, we obtain an average charge ratio of $Z_\sigma/Z = 0.23$ (Fig. S1 in the SI) and an average value for the bare charge of $Z = (10.4 \pm 0.5) \times 10^3$.

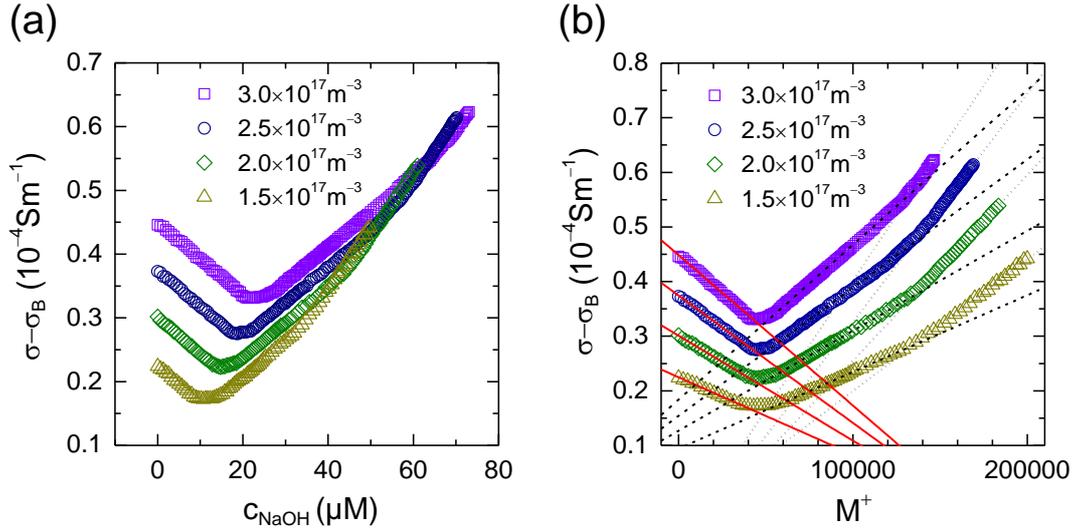

**Figure 2. $CO_2$-free conductometric titration with NaOH.** (a) Background-corrected conductivities in dependence on the concentration of added NaOH. Data are taken for different number densities as indicated in the key. (b) The same, now plotted in dependence on the number of cations added per particle. Dashed and dotted black lines denote fits of linear functions to the middle and final part of the curves, respectively. From their crossing point, we infer a total number of ionizable surface groups of $N_{CO2H} = (134 \pm 2) \times 10^3$. The red solid lines are least squares fits of Eqn (5) to the data. From these, we obtain an average value for the bare charge of $Z = (10.4 \pm 0.5) \times 10^3$.

*3.3 Addition of non-reacting electrolyte*

For the addition of 1:1 electrolytes to our suspensions of negatively charged colloidal spheres, we started from either a thoroughly deionized, salt and $CO_2$-free, or from a thoroughly deionized state subsequently equilibrated against ambient air. Figure 3 compares the background-corrected conductivities for different $n$ upon addition of a simple strong acid (HCl) and a simple electrolyte (NaCl). For HCl, we observe a linear increase irrespective of the state of carbonization (Fig. 3 a and b). For NaCl, the situation is qualitatively different in the absence and presence of $CO_2$. In the decarbonized state (Fig. 3c), the curves initially show a steep increase. They then bend over and continue in a linear fashion at lesser slope. With increasing $n$, the cross-over region between the two characteristic slopes shifts towards higher $c$. In the presence of $CO_2$, the dependence is again linear (Fig. 3d).

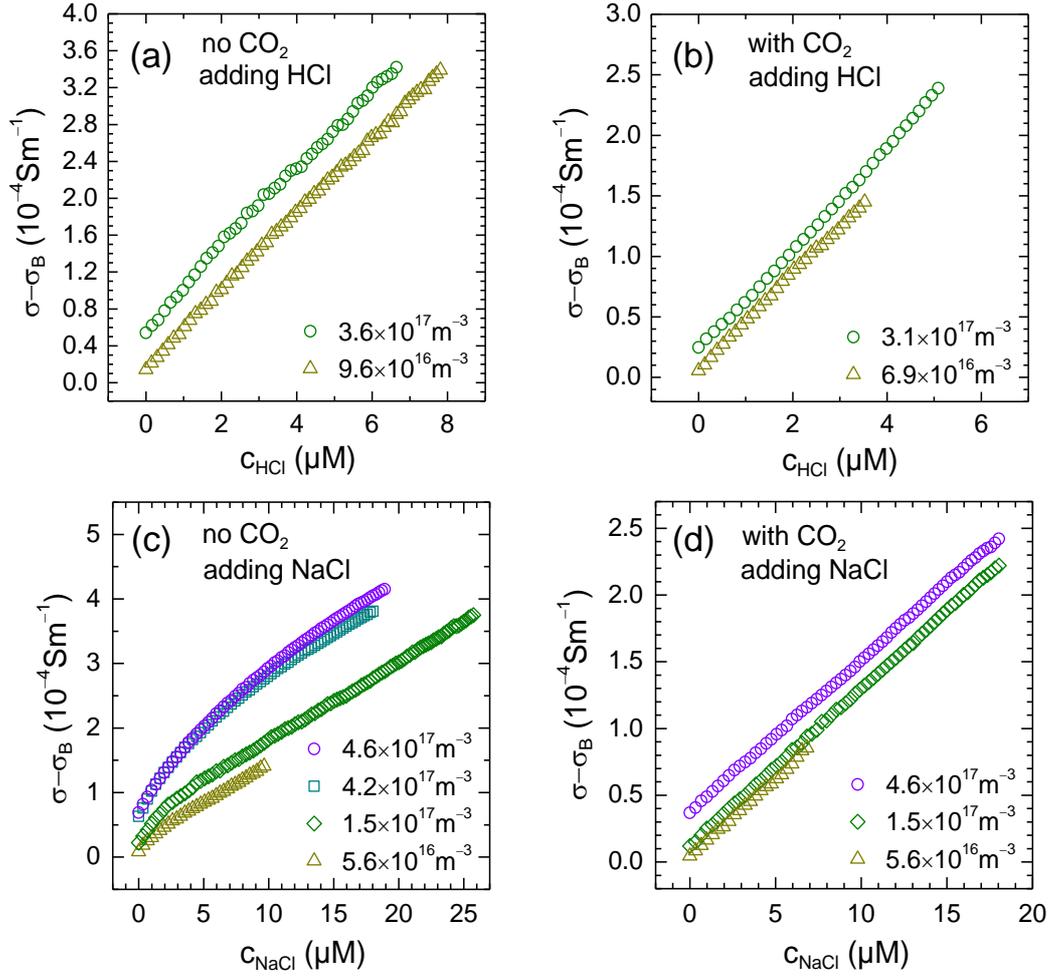

**Figure 3. Electrolyte concentration dependent conductivities of initially deionized suspensions.** Background corrected conductivities for $CO_2$-free (left) and $CO_2$-equilibrated (right) conditions and different number densities as indicated in the key. (a) and (b) Addition of a monoprotic strong acid: HCl. (c) and (d) Addition of a monovalent simple electrolyte: NaCl.

For each situation, the data collapse on single master curves, when scaled by the particle number density and plotted versus number of added cations per particle, $M$ (Fig. S2 and S3 in the SI). To isolate the conductivity contribution of added electrolyte per particle, $(\sigma - \sigma_0 - \sigma_B)/n$, data are additionally corrected for the particle contribution to conductivity, $\sigma_0$ (Fig. S4 and S5 in the SI). In Fig. 4, we show the collapsed data for the addition of NaCl to $CO_2$-free and to $CO_2$-equilibrated suspensions. Fig. 4a and c focus on the initial stages of the titrations and emphasize the different shapes of the master curves. Without $CO_2$, the addition of small amounts of NaCl yields a bent curve initially following the limiting conductivity contribution of HCl. At larger $M$, the curve straightens and runs parallel to the limiting conductivity contribution of NaCl. In the presence of $CO_2$, we obtain a linear increase with a slope corresponding to the limiting conductivity contribution of NaCl. Figs. 4b and d highlight the collapse of data taken at different $n$ over several orders of magnitude in salt concentration. In the absence of $CO_2$ we see a well-defined transition between the two linear regimes with its turning point occurring around $M \approx 1 \times 10^4$.

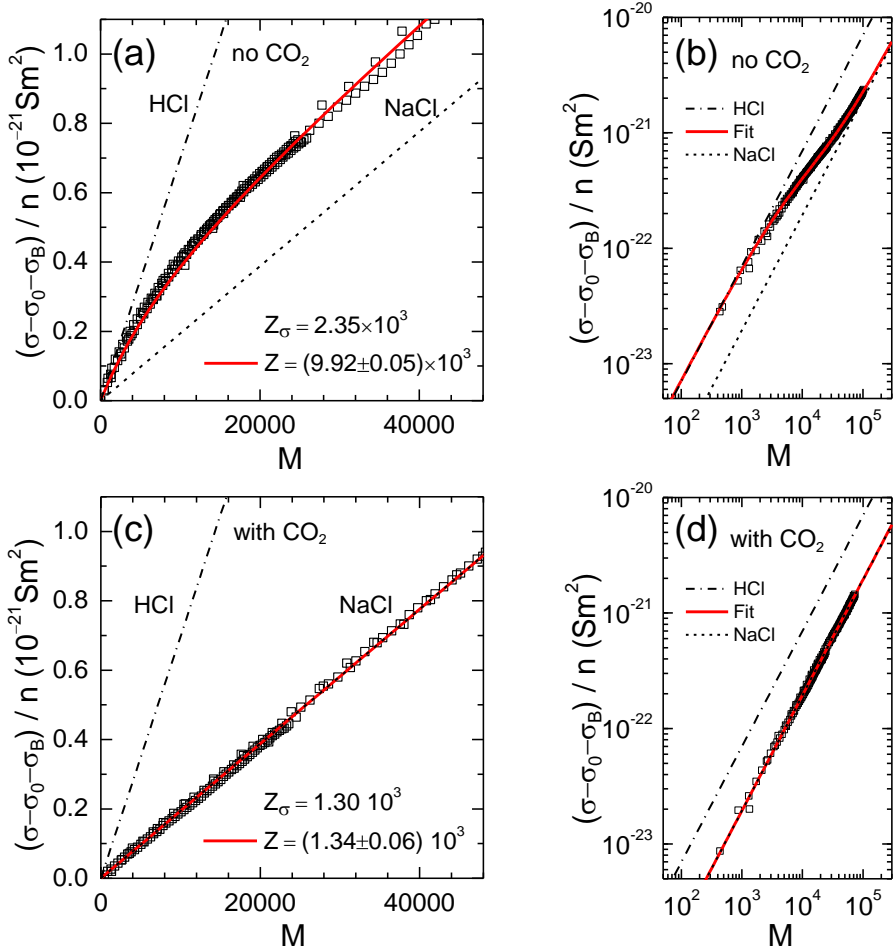

**Figure 4. Conductivity contribution of added electrolytes per particle under ideal (no $CO_2$) and environmental conditions (with $CO_2$).** The data are plotted in dependence on the number of $Na^+$ added per particle, $M$. The red solid lines are least squares fits of Eqn (6) to obtain the corresponding particle bare charge, $Z$, as indicated in the keys. For comparison, we also show the limiting conductivity contributions HCl (das-dotted) and NaCl (short dashed). (a) No $CO_2$. The fit of Eqn (6) returns $Z = (9.92\pm0.05)\times10^3$, with the error denoting the standard error at a confidence level of 0.95. (b) Double logarithmic rendering of the same data. A transition between two linear regimes is observed at $M \approx 1\times10^4$. (c) and (d) With $CO_2$. The fit of Eqn (6) now returns $Z = (1.34\pm0.06)\times10^3$.

To describe the data quantitatively, we rewrite Eqn (2) to read:

$$(\sigma-\sigma_0-\sigma_B)/n = e\left(Z_\sigma\left(\langle\mu^+\rangle-\mu_{H^+}\right) + M\left(\langle\mu^+\rangle+\mu_{Cl^-}\right)\right) \quad (6)$$

Note that with $Z_\sigma$ known from the previous experiment, the only free parameter in Eqn (6) is the charge ratio $Z_\sigma/Z$, respectively the bare charge $Z$ entering via <μ⁺> (see also Fig. S8 of the SI). Fits of Eqn (6) to the data use the conductivity charges $Z_\sigma = (2.35 \pm 0.04)\times10^3$ and $Z_\sigma = (1.30 \pm 0.07)\times10^3$ as input for the $CO_2$-free and the $CO_2$-saturated state, respectively. They return $Z = (9.92 \pm 0.05)\times10^3$ and $Z = (1.34 \pm 0.06)\times10^3$. In the absence of $CO_2$, the derived bare charge is in good agreement with the result from the titrations with NaOH. However, the presence of $CO_2$ reduces the bare charge to about 14% of its value under $CO_2$-free conditions.

Within Hessinger´s model, our results signify the exchange of low-mobility $Na^+$ added to the mobile part of the EDL for highly mobile $H^+$ stemming from the stagnant part of the EDL [58]. Initially, practically all $Na^+$ are exchanged for protons and the bulk conductivity increases strongly with a slope similar to the limiting conductivity contribution of HCl. The increase slows as the compositions in the mobile and stagnant parts of the EDL approach each other. After completion of the exchange, the slope is determined solely by the addition of NaCl to water. The situation is strikingly different in the $CO_2$-saturated case. Here, any curve bending effect is practically absent. This signifies the absence of any ion exchange between the mobile and the immobile fractions of the micro-ionic population. This can be rationalized considering the here observed coincidence of $Z_\sigma$ and $Z$. The total number of freely mobile counter-ions coincides with the total number of counterions: practically all counter-ions are freely mobile at this low bare charge. The same behavior is seen for the addition of KCl and CsCl. This shows that the presence of exchange in the $CO_2$-free state and its absence in the $CO_2$-saturated state are well reproducible and independent of the type of cation used (Fig. S7 of the SI).

The absence of exchange seen in Fig. 4c and d is different to its non-observability upon the addition of HCl (cf. Fig. S4a and b and Fig. S6 of the SI) There, we have an identity of counter-ions and added cations. No change in average cation mobility occurs and thus no change of slope.

**4. Charge Renormalization of Charge-Regulating Colloids – Poisson-Boltzmann Theory**

To rationalize our experimental observations, we utilize Poisson-Boltzmann theory. We employ the cell-model approximation [66] to reduce the complicated many-body system of a colloidal solution to a computationally tractable problem. Thus, we consider a single spherical colloid of radius $a$, centered in a spherical cell of radius $R = a\,\Phi^{-1/3}$ that is overall electroneutral. For computational convenience, we couple the cell at its outside to a reservoir at a given pH-value. Note that the reservoir is a purely theoretical construct which has no analogue in the experimental setup. Due to the spherical symmetry, the problem reduces to a one-dimensional problem in the radial coordinate $r \in [a, R]$. In order to obtain the (reduced) electrostatic potential $\Psi$ and the distribution of ions, we solve the Poisson-Boltzmann equation

$$\left(\frac{d^2}{dr^2} + \frac{2}{r}\frac{d}{dr}\right)\Psi(r) = \kappa_0^2 \sinh(\Psi(r)), \tag{7}$$

where $\kappa_0 \equiv \sqrt{8\pi\lambda_B n_0}$ is the inverse Debye length of the reservoir with $\lambda_B = e^2/(4\pi\varepsilon_0\varepsilon_r k_B T)$ denoting the Bjerrum length. In the salt-free (apart from the autoionization) case considered here, $n_0$ and thus $\kappa_0$ is fully determined by the pH-value of the reservoir:

$$n_0 = \max\left(10^{-\text{pH}_{\text{res}}}, 10^{-(14-\text{pH}_{\text{res}})}\right) \text{mol L}^{-1} N_A. \tag{8}$$

The PB equation given in Eqn 7 is a second-order ODE that has to be supplemented with two boundary conditions. At the cell boundary, the electric field has to vanish due to the overall electroneutrality, resulting in the Neumann boundary condition

$$\Psi'(R) = 0. \tag{9}$$

At the surface of the colloidal particle, the condition

$$\Psi'(a) = \frac{-Z\lambda_B}{a^2} \tag{10}$$

has to hold, where $Z$ is the (dimensionless) charge of the colloid. However, due to the charge regulating nature of the surface groups, $Z$ is not fixed but also depends on the potential:

$$Z = -N\alpha\left(\text{pH}_{\text{surface}}, \text{pK}_a\right) = -N\alpha\left(\text{pH}_{\text{res}} + \log_{10}(e)\Psi(a), \text{pK}_a\right). \tag{11}$$

Here, $N$ is the total number of ionizable surface groups and $\alpha$ is the average degree of their ionization and e is Euler's number. The specific functional form of $\alpha$ depends on the employed model of charge regulation. In our calculations, we employ the Henderson-Hasselbalch equation

$$\alpha_{\text{HH}} = \frac{1}{1 + 10^{\text{pK}_a - \text{pH}_{\text{surface}}}}, \tag{12}$$

which is equivalent to the Langmuir isotherm [20]. Equation 10, Equation 11 and Equation 12 form the self-consistent charge regulation boundary condition of Ninham and Parsegian [67]. Note that the Langmuir isotherm in the Ninham-Parsegian theory neglects correlation effects between nearby sites on the colloidal surface [20,22]. Consequently, the pK$_a$-values reported in the following are understood to be "effective" pK$_a$-values that approximately account for the electrostatic interactions between the surface groups.

We use the boundary value problem solver implemented in the open-source library SciPy [68]. To solve the PB equation with the specified boundary conditions for a given value of pH$_{\text{res}}$ numerically, we use an iterative procedure, where the charge $Z$ of the colloid is adjusted until self-consistency is achieved. Because the colloidal particle cannot leave the cell, there emerges a Donnan potential $\Psi(R)$ between the reservoir and the cell, leading to a concentration difference of protons. In order to obtain the bare charge of the colloid for a desired bulk pH-value (which we identify with the local pH-value at the cell boundary, $pH_{bulk} = pH_{res} + \Psi(R) \log_{10}(e)$) we solve the PB equation for a range of reservoir pH-values, calculate the bulk pH for each of them and linearly interpolate the result to the desired bulk pH-value. Similarly, we obtain the renormalized/effective charge of the colloid, which we calculate from our numerical solution using the analytical prescription of Trizac et al. [17]. Note that there exist several distinct charge renormalization schemes, which result in slightly differently effective charges [61]. However, given the inherent uncertainty of the experimental system, we argue that the choice of the method is only of secondary importance and we opt for the most popular one. To ensure numerical stability of the solver, we start at a very low pH-value, where the colloid is barely charged, and increase the pH-value only in small steps. In each step, the numerical solution of the previous pH-value is used as initial guess, which leads to rapid convergence and avoids numerical instabilities. Fig. 5 displays the results obtained for the present particles of $a$ = 179.5 nm, $N$ = 134×10$^3$ and $\Phi$ = 4×10$^{-3}$.

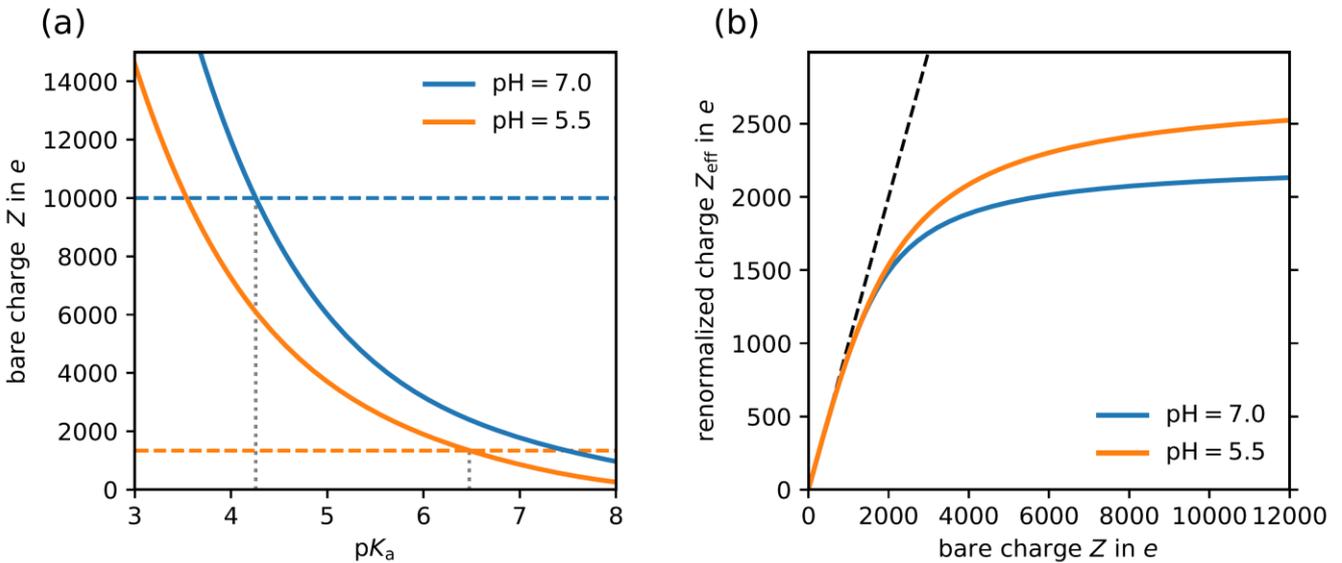

**Figure 5. Results from PBC-calculations.** Comparison of results for the two different situations indicated in the key: salt and $CO_2$-free state at $pH_{bulk}$ = 7 (blue) and $CO_2$-equilibrated state at $pH_{bulk}$ = 5.5 (orange). (a) Bare charges obtained for different surface group p$K_a$ (solid lines) to the experimentally observed bare charges (dashed horizontal lines). The crossing point yields surface p$K_a$ of 4.26 and 6.48, respectively. (b) Renormalized charge as a function of the bare charge. The black dashed line indicates the condition $Z_{PBC} = Z$.

In Figure 5a we show the calculated bare charge as a function of the surface-group p$K_a$-value for the two experimentally realized situations. The corresponding average bare charges from the experiments are indicated as dashed lines of the same color. In the $CO_2$-free state, we consider $pH_{bulk}$ = 7.0. Within the considered charge regulation model, we can identify a value of p$K_a$ ≈ 4.26 leading to the dissociation of $Z$ ≈ 10000 surface groups. In the $CO_2$-equilibrated state we consider $pH_{bulk}$ = 5.5 and $c_B$ = 3.16 µmolL$^{-1}$ and the experimentally determined bare charge is $Z$ ≈ 1340. We observe a drastic change in the surface-group p$K_a$ needed to realize this bare charge. In fact, the p$K_a$ shifts to an effective value of p$K_a \approx 6.48$.

The decrease of bare charge is mirrored by the accompanying decrease of the renormalized charge. Figure 5b shows the theoretically calculated renormalized charge as a function of the bare charge for both considered bulk pH-values. Initially following the relation $Z_{PBC} = Z$, it bends over and approaches a limiting value. Note that with $CO_2$, there are overall more ions in the cell, and, consistent with the results of Alexander et al. [15], the renormalized charge shows a larger saturation value. At pH$_{bulk}$ = 7.0, where the experimentally determined average bare charge was $Z = (10.4 \pm 0.5) \times 10^3$, the theory predicts a renormalized charge of $Z_{PBC} = 2100$. For pH$_{bulk}$ = 5.5, where the experimentally determined bare charge was $Z = 1340$, the theory predicts a renormalized charge of $Z_{PBC} = 1170$.

## 5. Discussion

We summarize our findings in Fig. 6. Figure 6a shows the results for the $CO_2$-free cases. The average values of $N = (134 \pm 2) \times 10^3$, $Z = (10.4 \pm 0.5) \times 10^3$ and $Z_\sigma = (2.35 \pm 0.04) \times 10^3$ reflect the well-known sequence of charge number magnitudes: $N >> Z > Z_\sigma$. Only a very small part of the weakly acidic surface groups is dissociated and only a fraction of the total number of counter-ions is actually mobile. From the PBC-calculations with pH-driven charge regulation (Fig. 5a) we infer a corresponding p$K_a$ of 4.26. This is slightly lower than previously reported values for carboxylate stabilized particles [3]. However, it appears well-compatible with a dominance of carboxylate surface groups and a small admixture of a few sulfate groups stemming from the initiator. The calculations further predict a renormalized charge of $Z_{PBC} = 2.1 \times 10^3$. This figure is in reasonable agreement with the experimentally obtained value. The small deviation is attributed to the simplifications made in the present theoretical approach, e.g., the neglect of microscopic surface geometry and of ion correlations. Overall, the comparison demonstrates the excellent performance of Hessinger´s conductivity model.

In Fig. 6b, we replot the $Z$- and $Z_\sigma$-data from the $CO_2$-free experiments on an enlarged scale and compare them to the results obtained in the presence of $CO_2$ (reddish symbols). For the latter, we found an average bare charge of $Z = (1.34 \pm 0.06) \times 10^3$ and an average conductivity charge of $Z_\sigma = (1.30 \pm 0.06) \times 10^3$. If $CO_2$ had no influence on the p$K_a$, we would expect a bare charge of $Z \approx 6 \times 10^3$. However, the much smaller experimental bare charge is observed in the charge regulating PBC-calculations for an effective surface p$K_a$ of p$K_a \approx 6.48$. The PBC-prediction for the renormalized charge at pH = 5.5 then is $Z_{PBC} = 1200$ in reasonable agreement with the experimentally obtained value (The small deviation is, again, attributed to the neglect of ion correlations). At such low bare charge numbers, there is no substantial charge renormalization. If we follow previous work and identify $Z_{PBC}$ with $Z_\sigma$, this finding explains the absence of ion exchange in the presence of $CO_2$. There are, simply speaking, hardly any immobile ions to be exchanged. The final charge number magnitude sequence reads: $Z_\sigma^{+CO_2} \approx Z^{+CO_2} < Z_\sigma^{-CO_2} << Z^{-CO_2} << N$.

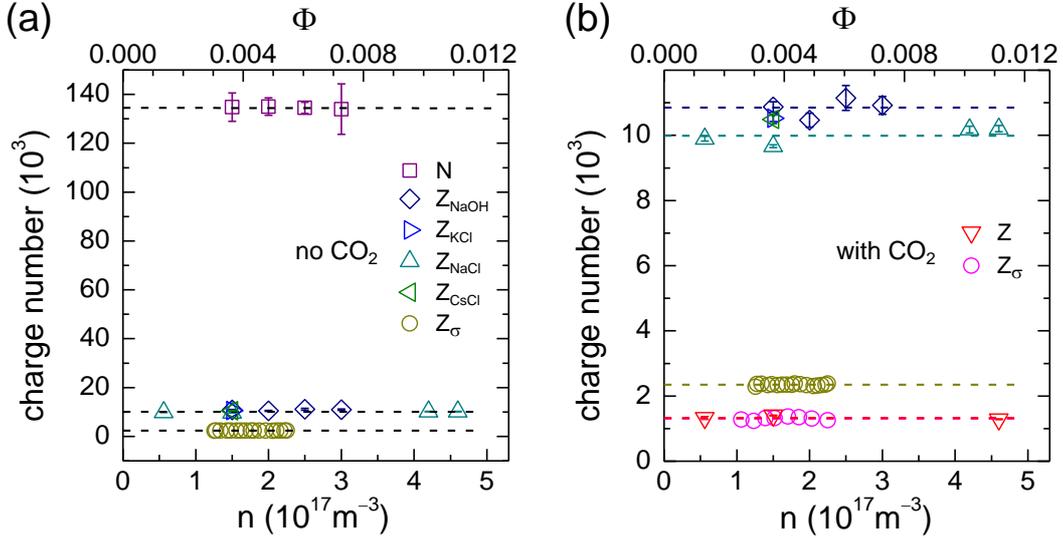

**Figure 6. Comparison of charge numbers for different particle number densities.** Here we display the individual fit results obtained at different densities for the different experiments (symbols as indicated in the keys) and the corresponding averages thereof (dashed lines). The upper scale represents the corresponding particle volume fraction, $\Phi = (4/3)\pi a^3 n$. (a) Comparison of different types of charge for the salt- and $CO_2$-free systems. We obtained the following averages with quoted errors denoting the standard deviation: number of dissociable surface groups $N = (134 \pm 2) \times 10^3$, average bare charge number from titration with NaOH: $Z = (10.7 \pm 0.5) \times 10^3$, average bare charge number from titration with NaCl: $Z = (10.0 \pm 0.3) \times 10^3$, barge charge number from titration with KCl: $Z = (10.5 \pm 0.1) \times 10^3$, bare charge number from titration with CsCl: $Z = (10.5 \pm 0.1) \times 10^3$, average effective conductivity charge $Z_\sigma = (2.35 \pm 0.04) \times 10^3$. (b) Comparison of bare and effective charges for $CO_2$-free systems (replotted from (a)) and $CO_2$-equilibrated systems on an enlarge scale. For the $CO_2$-equilibrated systems, we find the average bare charge number from titration with NaCl, $Z = (1.34 \pm 0.04) \times 10^3$ and the average effective conductivity charge, $Z_\sigma = (1.30 \pm 0.07) \times 10^3$.

The drastic reduction of the effective conductivity charge and the renormalized charge in the presence of $CO_2$ is in line with the previous electrophoretic results [55,57], where we had reported a decrease of the $\zeta$-potential from $\zeta = (61 \pm 4)$ mV to $\zeta = (35 \pm 2)$ mV after equilibration to ambient air and even further to $\zeta = (3 \pm 2)$ mV after equilibration to a $CO_2$ atmosphere. These experiments could, however, not directly demonstrate an underlying decrease of the bare surface charge. The present combination of conductivity experiments under various conditions with charge-regulating PBC-calculations unequivocally revealed that $CO_2$ directly exerts a strong decharging effect on the bare surface charge. The reduction in $\zeta$-potential and conductivity charge, respectively renormalized charge simply follows suit, and the absence of ion exchange effects upon the addition of pH neutral 1:1 electrolytes to $CO_2$-equilibrated systems is readily explained by the negligible amount of charge renormalization under these conditions.

Granting that the observed effect can be expressed as a significant shift of the surface $pK_a$, the observed shift is much larger than the known difference for COOH-groups in the bulk and at surfaces, which typically amounts to $\approx 0.5$ $pK_a$ units. Further the observed shift by more than two units exceeds by far the expectable uncertainties in the theoretical calculations introduced by employing the PBC model. We therefore believe that this unforeseen observation poses a considerable challenge to explain, how dissolved molecular $CO_2$ could possibly influence the dissociation equilibrium. In our previous work, we had tentatively considered several alternatives regarding the underlying mechanisms. All of them were motivated by the results of our molecular dynamics simulations, where we observed the formation of a diffusely adsorbed monolayer of $CO_2$ very close to the surface [55]. In principle, such a layer could lead to surface modifications shifting the shear plane further outward and thus decrease the $\zeta$-potential without influencing the particle charge [42,69]. Since we here directly observed a decrease in the bare charge, we can now rule out this possibility. Alternatively, the layer could decrease the dielectric permittivity close to the surface and thus strengthen any electrostatic interactions. Enhanced attractive interactions could increase the dissociation energy. Stronger repulsion between neighboring surface charges could introduce a Coulomb penalty for dissociation. Both would counteract dissociation and the corresponding shift in effective surface $pK_a$ would directly lower the bare charge. Which of these effects dominates the observed dielectric charge regulation cannot be decided within the present approach. A detailed treatment remains a theoretical challenge but should be accessible within recent analytic approaches accounting for spatially variable dielectric permittivities, hydrodynamics and finite ion size effects [32,70]. It should also be in reach of computer simulations [71]. Overall, we anticipate, that the present observations should motivate enhanced theoretical interest in an improved modelling of double layer properties.

On the experimental side, the $CO_2$ layer could be characterized further by X-ray and/or element selective neutron reflectometry [40]. Its consequences for colloid-colloid interactions could be determined with high distance resolution by colloidal probe experiments [2,10,11]. But also the conductivity experiments could be extended to include a back-titration by HCl starting from the NaCl-saturated state. Up to now, we had studied polymer latices stabilized by weak acids, hydrophilic silica surfaces, and super-hydrophobically coated surfaces, but also further surfaces should be addressed to firmly establish the $CO_2$ adsorption as general phenomenon.

A final remark addresses the implications of our findings for future experiments at low salt conditions. First, the significant influence of $CO_2$ on the charge state of dielectric surfaces in contact with aqueous electrolytes will require utmost care in characterizing and controlling its concentration in any precision measurement of suspension properties and surface characteristics. The techniques employed in the present paper may serve as a starting point. Second, the novel decharging effects exerted by molecular $CO_2$ will certainly affect many surface and suspension properties. Occasional consequences of $CO_2$-contaminations have been reported [51]. Möller et al. used treatment of surfaces with $CO_2$-equilibrated DI water and subsequent partial $CO_2$-removal to improve improved control for writing and drawing in water via modified diffusiophoretic transport of the ink particles [72]. Systematic experiments comparing the struc-ture and phase behavior in dependence on electrolyte concentration and number density with and without $CO_2$ are now under way and will be reported elsewhere. Overall, we believe that the controlled application of molecular $CO_2$ will open novel ways of manipulating the performance of functional colloidal systems and solid-water interfaces.

## 5. Conclusions

We demonstrated the direct influence of molecularly dissolved $CO_2$ on the state of surface group dissociation. We observe a drastic reduction of the number of dissociated surface groups and a halving of effective charges. Such a reduction is significantly stronger than the expected effects induced by the $CO_2$-reaction products in salt-free water, i.e., an altered local pH and an increased salinity caused by the formation of carbonic acid which partially dissociates. The present study therefore confirmed and substantially extended previous investigations and further supported our suggestion of an additional charge regulation by $CO_2$. Given the relatively small concentration of $CO_2$ in ambient air of 421 ppm, the extent of decharging dielectric surfaces by a charge neutral molecule is equally impressive and surprising. In fact, while there existed some literature on the effects of degassing in general, the specific effects of $CO_2$ had been largely ignored. However, in view of the ubiquitous nature of water-solid interfaces and their importance in geology, meteorology, biology as well as in technical applications, we feel that much further research is needed to explain our unexpected findings.

**Acknowledgements**

We thank J. Yamanaka, L. Belloni and Y. Levin for fruitful discussions. Financial Support by the German Research Foundation (DFG) is gratefully acknowledged. D. B. and C. H. received funding within the Research Unit FOR2811 "Adaptive Polymer Gels with Model Network Structure" under grant 423791428 along with grant 397384169 (TP7). C. H. is partially funded by the Deutsche Forschungsgemeinschaft (DFG) in project ID 358283783 – CRC 1333/2. P. V. and T. P. were financially supported by the DFG through grant nos. Pa459-18.2, Pa459-19.1 and Pa459-23-1.

**Data availability statement**

Raw data are obtainable from the authors upon reasonable request. The source code used for the PB cell-model calculations is freely available on GitHub (https://github.com/davidbbeyer/pb_solver).

**Author Contributions**

P. V. performed the experiments and the evaluations. D. B. performed the PBC-calculations. C. H. supervised the theoretical and T. P. the experimental part. P. V. and T. P. wrote the original manuscript. All authors contributed equally in discussing the data and finalizing the manuscript.

**Entry for the Table of Contents**

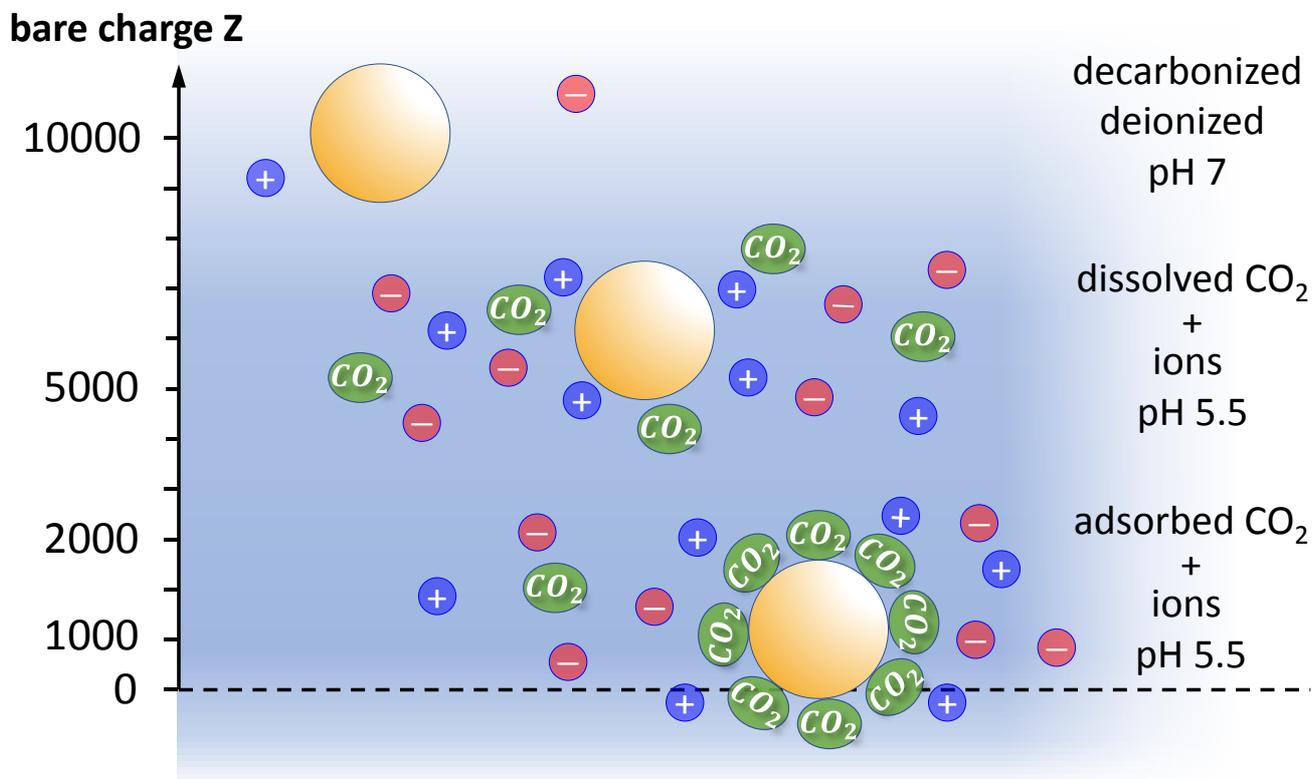

**TOC caption:** Under deionized and decarbonized conditions, the sphere is highly charged. Expected screening- and pH-effects lower the charge. Diffusely adsorbed $CO_2$ causes drastic decharging.

**Keywords:** charged colloids • electrical double layer • conductivity • $CO_2$ adsorption

# Supporting information

# for

# $CO_2$-induced Drastic Decharging of Dielectric Surfaces in Aqueous Suspensions

Peter Vogel[1]*, David Beyer[2], Christian Holm[2], and Thomas Palberg[1]

[1] Institute of Physics, Johannes Gutenberg University, 55128 Mainz, Germany

[2] Institut für Computerphysik (ICP), Universität Stuttgart, 70569 Stuttgart, Germany

* corresponding author

## 1. Additional Data from evaluation.

In Fig. S1 we show the results for the charge ratio as obtained from the fits of Eqn (5) to the NaOH titration data in the limit of small amounts of added NaOH. The error bars represent the standard error of the fits at a confidence level of 0.95 and are smaller than the symbol size. The charge ratio $Z_\sigma/Z$ as averaged over these four experiments is 0.23±0.01, where the error now represents the standard deviation. Taking the number of acidic surface groups, $N = 134 \times 10^3$, we obtain a degree of dissociation of $\alpha_{eff} = Z/N = 0.018$.

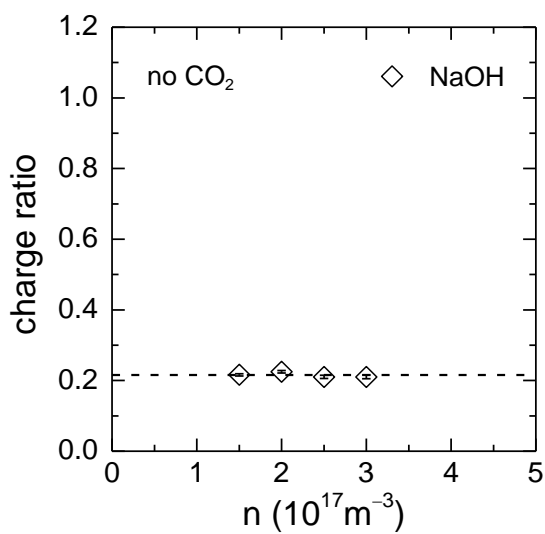

**Figure S1 charge ratio $Z_\sigma/Z$ obtained from conductivity titration.** Error bars represent the standard errors of Fits of Eqn (5) to the data at low values of added base. The dashed line denotes the average value of $Z_\sigma/Z$ = 0.23±0.01.

Figures S2 to S4 give a step by step illustration of our data treatment. In Fig. S2a to d, we show the data of Fig. 4 replotted on adjusted scales in dependence on the number of micro-ions added per particle, $M$. The resulting slopes show a significant dependence on the number density $n$. Moreover, the difference between initial and final slopes gets enhanced in this drawing.

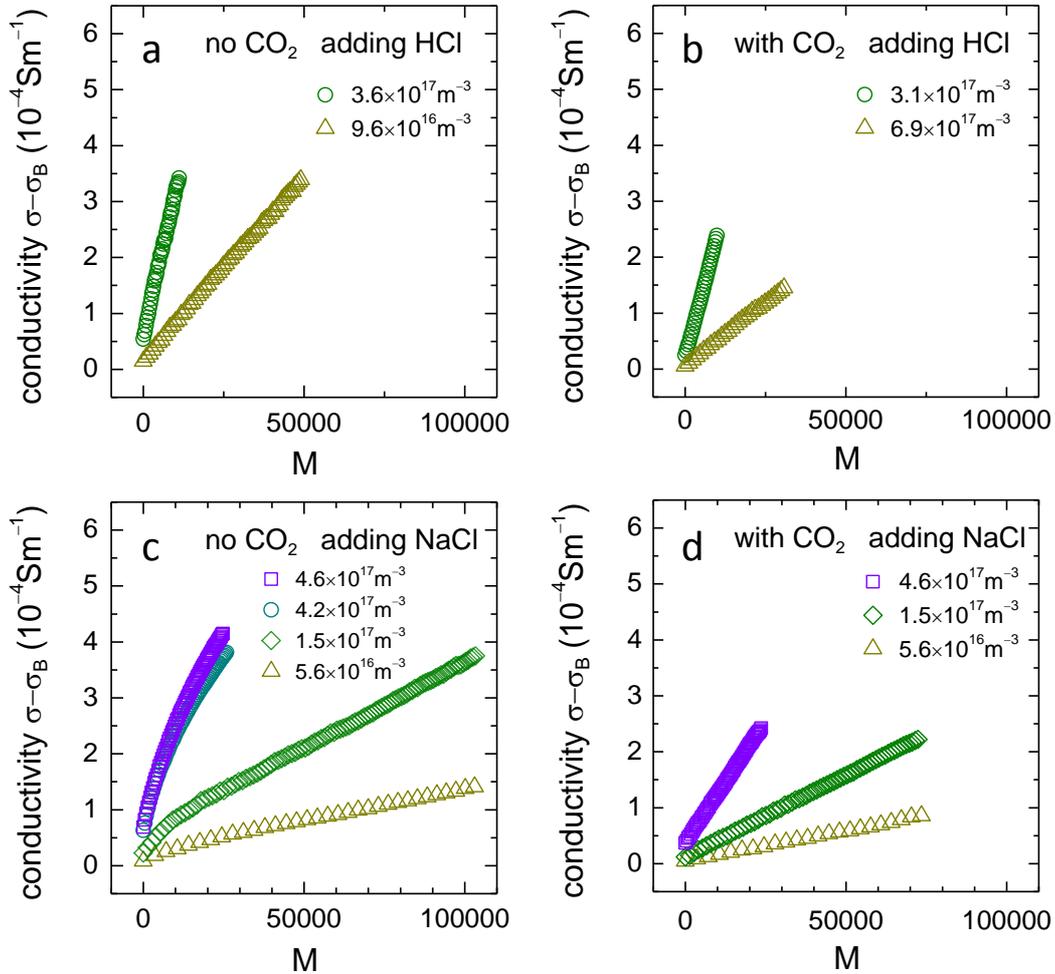

**Figure S2 Background-corrected conductivity in dependence on added electrolyte per particle.** Starting conditions and type of added electrolyte are indicated in each key.

In Fig. S3 we show the data for all four experiments after normalizing the conductivity by the particle number density. The semi-logarithmic plot of Fig. S3 shows the collapse onto single master curves over several orders of magnitude. In the limit $M \to 0$, data points approach constant values related to the respective particle effective charge, $Z_\sigma$, in absence or presence of $CO_2$.

The background and particle contribution corrected conductivity normalized by the particle density is shown in Fig. S4 in dependence on $M$. This isolates the conductivity contribution of the electrolyte per particle. Fig. S4 again highlights the differences in curve shape and clearly shows the presence of ion exchange only in the case of adding NaCl to a deionized and decarbonized system (Fig. S4c). In all other cases the conductivity increases linearly with *M*. Note that in Fig. S4b, one observes a marginally significant deviation from the expected slope, becoming visible for the largest amounts of added HCl.

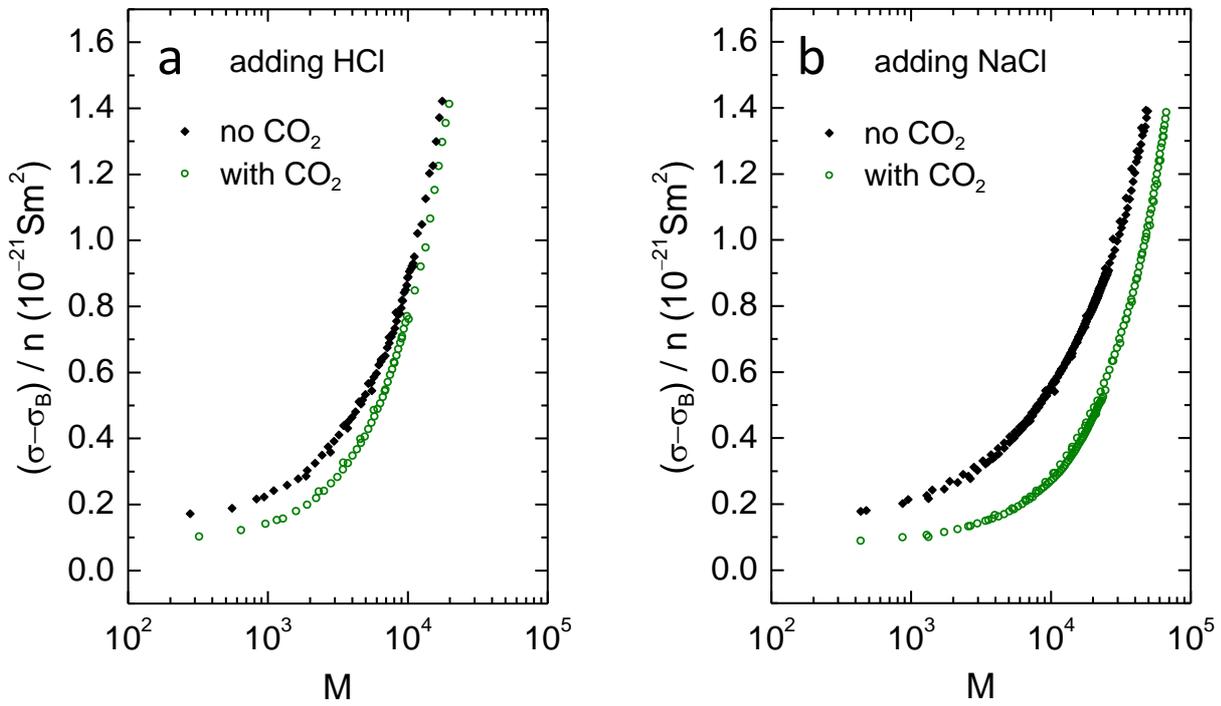

**Figure S3 Background-corrected conductivity per particle in dependence on added electrolyte per particle**. Starting conditions and type of added electrolyte are indicated in each key. For each experiment, the conductivities collapse on a single master curve over several orders of magnitude in *M*. Under ambient conditions (with $CO_2$), the conductivity contribution per particle is significantly lowered.

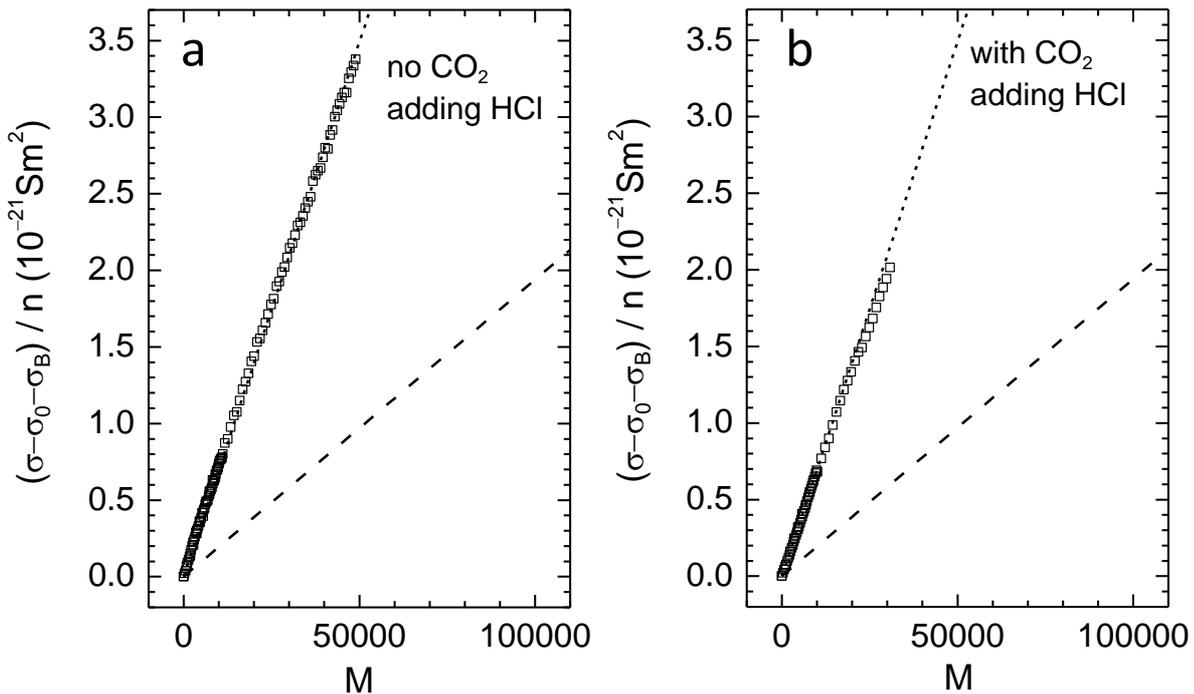

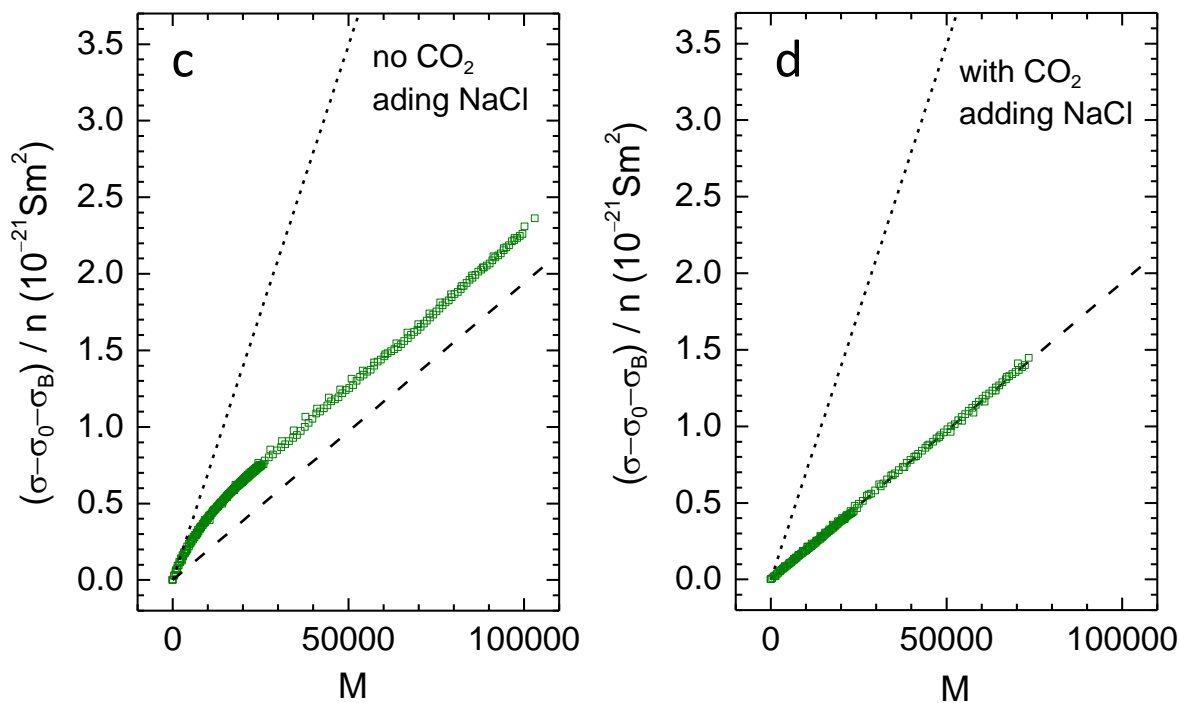

**Figure S4. Conductivity contribution of added electrolytes per particle under different conditions**. Data are plotted in dependence on the number of added electrolyte per particle. Starting conditions and type of added electrolyte are indicated in each key. For comparison, the dotted and dashed lines represent the limiting conductivity contribution of HCl and NaCl, respectively. (a) Addition of HCl in the absence of $CO_2$. (b) Addition of HCl in the presence of $CO_2$. (c) Addition of NaCl in the absence of $CO_2$. (d) Addition of NaCl in the presence of $CO_2$.

In Fig. S5, we replot the data of Fig. 4 in a double logarithmic fashion. The data for the addition of NaCl to a deionized and decarbonized suspension switch between the two limiting linear behaviors. The initial exchange of $H^+$ from the diffuse part of the EDL for $Na^+$ from the stagnant layer is nearly quantitative. It slows in the transition region, and for large $M$, it has practically ceased. In all other cases, a global fit of a power law with exponent 1 describes the data well. (The small deviation from linearity seen before in Fig. S4b is not discernible in this rendering).

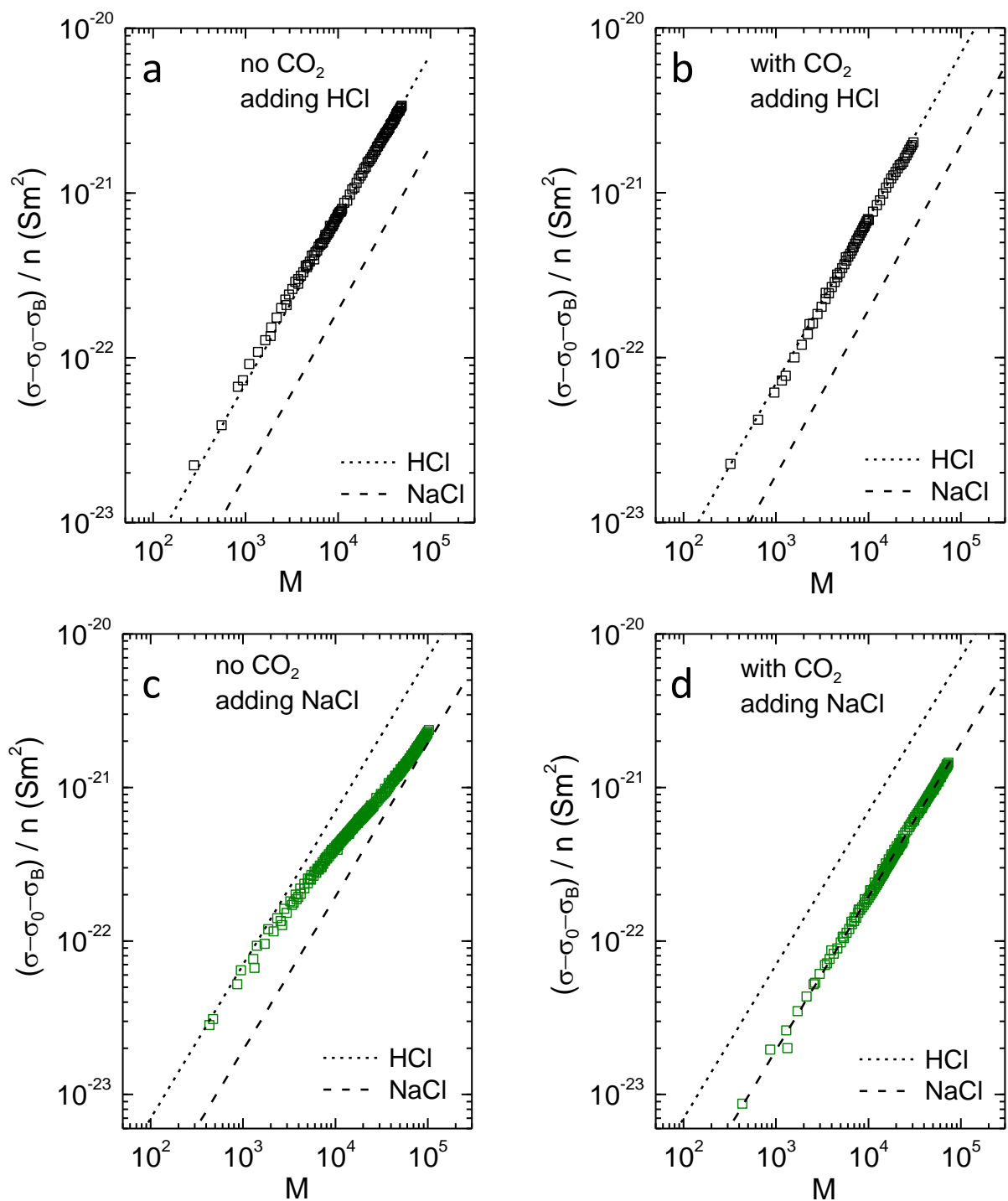

**Figure S5. Conductivity contribution per particle in dependence on the number of 1:1 electrolyte ions added per particle**. Data are plotted in a double logarithmic fashion. Starting conditions and type of added electrolyte are indicated in each key. The dotted and dashed lines represent the conductivity contribution of pure HCl and NaCl, respectively. (a) Addition of a HCl in the absence of $CO_2$. (b) Addition of HCl in the presence of $CO_2$. (c) Addition of NaCl in the absence of $CO_2$. (d) Addition of NaCl in the presence of $CO_2$.

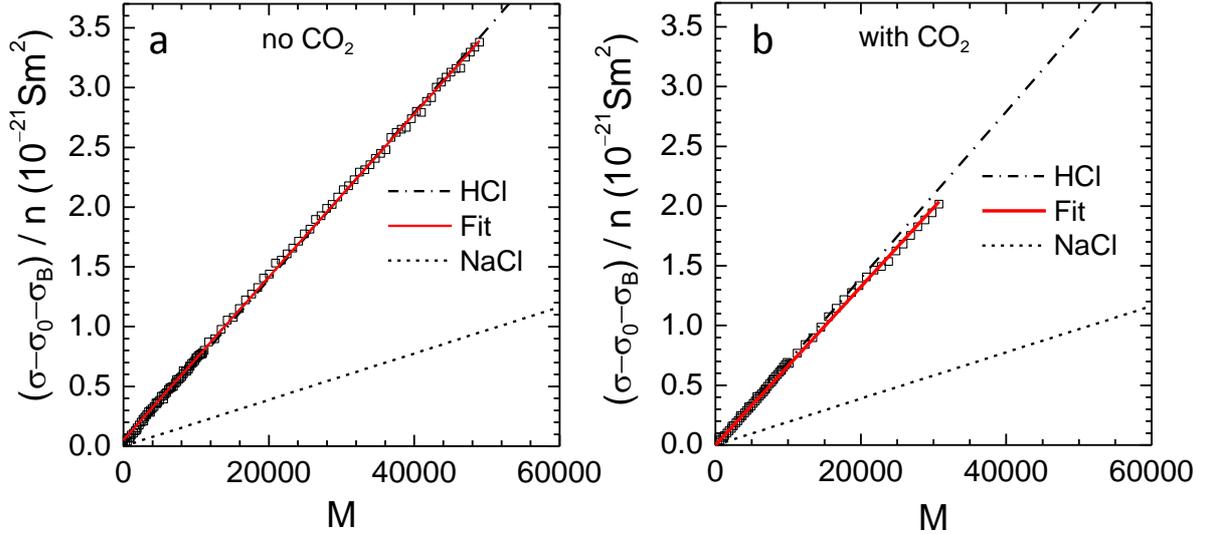

**Figure S6. Conductivity contribution of an added simple acid under ideal (no $CO_2$) and environmental conditions (with $CO_2$).** Also shown are least squares fits linear functions (red lines) do the data. As before, the dotted and dashed lines represent the conductivity contribution of pure HCl and NaCl, respectively. (a) Decarbonized conditions. Data are excellently described by the fit of Eqn (6) coinciding with the expectation for the conductivity contribution of HCl. (b) $CO_2$-saturated conditions. Within experimental scatter, data can again be described by a linear function, which, however does not coincide with the expectation for the conductivity contribution of HCl.

In the main text, we had compared the conductivity contribution of added NaCl to fits of Eqn (6). In Fig. S6, we show the corresponding data for the addition of HCl. Both data sets are excellently described by linear functions, signifying the absence of ion exchange. In fact, no exchange is anticipated for the addition of HCl, as the anion Cl⁻ anyway stays in the mobile part of the EDL, while the added cation H⁺ is identical to the ion species of the stagnant part. Thus, $\langle\mu^+\rangle = \mu_{H+}$, and there is no change in the average micro-ion mobility. Therefore, the slope in Fig. S6a coincides with that of the limiting conductivity contribution of HCl. In Fig. S6b, the slope falls slightly below this expectation. This can be attributed to additional pH driven charge regulation leading to a slight further reduction of $Z$ and thus of $Z_\sigma$. A decrease of by pH driven charge regulation should also be present for the addition of HCl to a decarbonized system (Fig. S4a) but cannot not show up here, as $Z_\sigma$ is already in the saturation regime.

The experiments were repeated using KCl and CsCl as added electrolytes to check for any influences of cation type on the exchange process. The corresponding data are shown together with Fits of Eqn (6) in Fig. S7. Using the conductivity charge of $Z_\sigma = 2350$ as input we obtain bare charges of $Z = (10.52 \pm 0.10) \times 10^3$ for KCl and $Z = (10.49 \pm 0.14) \times 10^3$ for CsCl. The double logarithmic data renderings show a transition between the two linear regimes with the turning point occurring around $M \approx 1 \times 10^4$.

Therefore, the data show exactly the same behaviour as seen for the addition of NaCl, and any influence of the type of cation can be excluded.

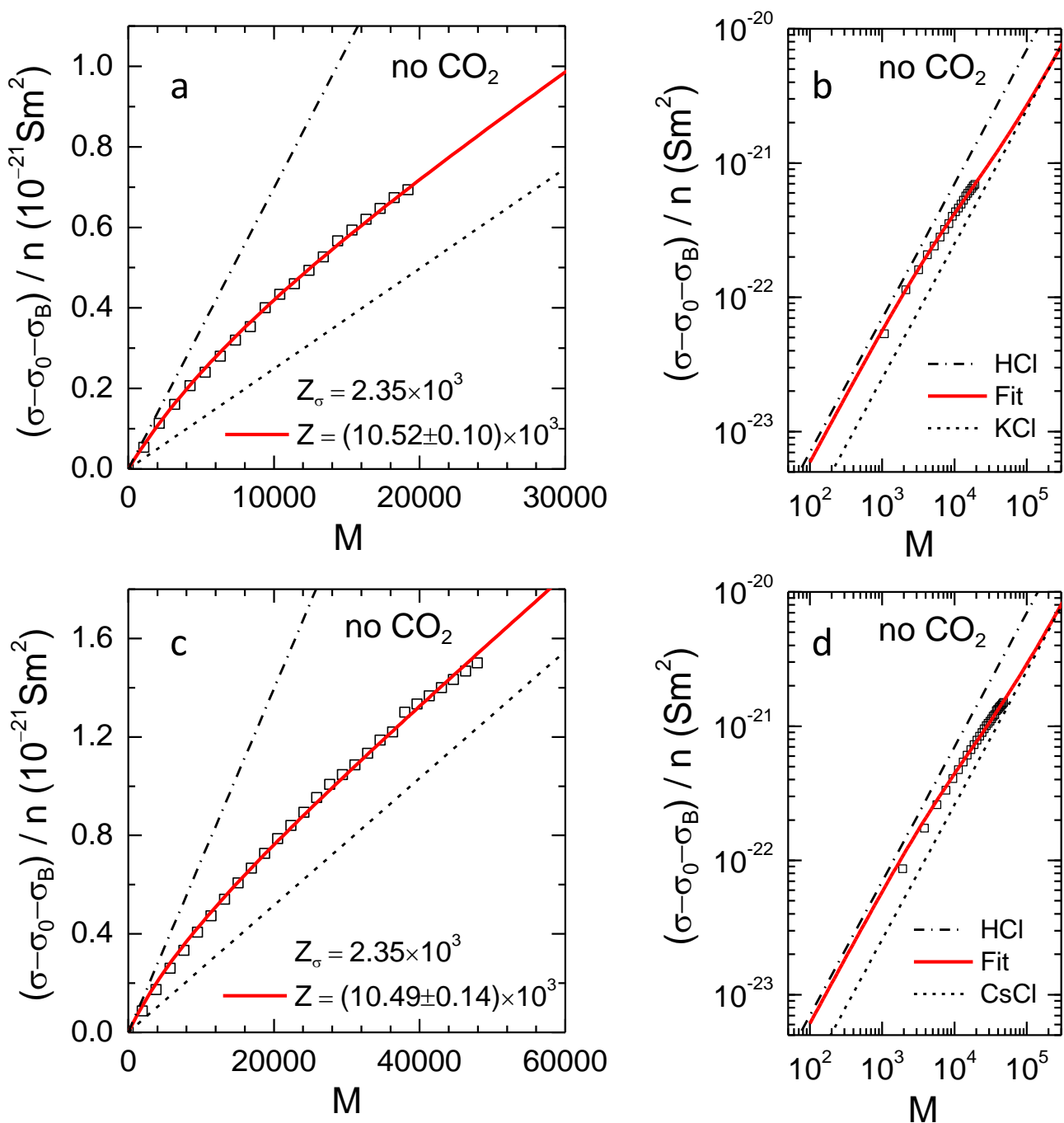

**Figure S7. Conductivity contribution of added electrolytes under CO$_2$-free conditions for different cation types.** The upper row shows data for the addition of KCl, the lower row shows data for the addition of CsCl. The data are plotted in dependence on the number of micro-ions added per particle. The red solid lines are least squares fits of Eqn (6) to obtain the corresponding particle bare charge, $Z$, as indicated in the keys. The dash-dotted and short dashed lines represent the limiting conductivity contribution of HCl, and of KCl and CsCl, respectively. (a) and (c) show linear, (b) and (d) show double logarithmic renderings of the data.

## 2. Dependence of Eqn. (6) on the charge ratio

The behaviour of Eqn. (6) is illustrated in Fig. S8 for selected charge ratios $Z_\sigma/Z$ and $\sigma_B = 0$. Note that $\sigma$ depends on $Z$ entering via $\langle \mu^+ \rangle$. The first part of the electrolyte contribution goes to zero whenever $\langle \mu^+ \rangle$ equals $\mu_{H^+}$. Our calculations show this to be the case in the limit of $M \ll Z$ and in particular at small charge ratio $Z_\sigma/Z$. Thus, the initial slope in a plot of $(\sigma-\sigma_0)/n$ versus $M$ will correspond to the limiting conductivity contribution of HCl (dash-dotted line). Upon further addition of neutral electrolyte, the number of added $Na^+$ exchanged for $H^+$ decreases successively, while the number of counter-ions, $Z$, is conserved. In the limit of $M \gg Z$, practically all immobilized $H^+$ have been exchanged for $Na^+$, ion exchange has ceased, and the slope approaches that of the limiting conductivity contribution of NaCl (short dashed line). The experimentally determined charge ratio was 0.23 corresponding roughly to the blue curve.

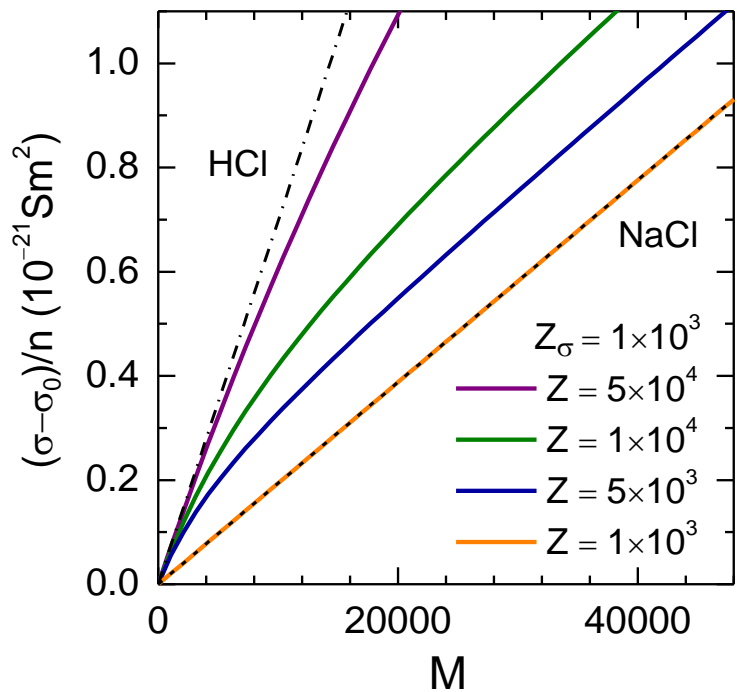

**Figure S8. Conductivity contribution of added electrolytes.** Model calculations using Equation (6) show the conductivity contribution per particle as function of the number small ions added per particle, $M$. The results for fixed effective charge and different bare charges are shown as color-coded solid lines, as indicated in the key. From top to bottom, the corresponding charge ratios are $Z_\sigma/Z = 0.02, 0.1, 0.2, 1$. We also show the two limiting linear regimes reached in systems with vanishing exchange. The limiting conductivity contribution of HCl (dashed-dotted line) is met for the addition of HCl irrespective of $Z$ as long as $Z \gg Z_\sigma$, as well as for addition of NaCl for $Z = \infty$. The limiting conductivity contribution of NaCl (dashed line) is met for addition of NaCl for $Z = Z_\sigma$.